\documentclass[fleqn,usenatbib]{mnras}

\usepackage{newtxtext,newtxmath}

\usepackage[T1]{fontenc}

\DeclareRobustCommand{\VAN}[3]{#2}
\let\VANthebibliography\thebibliography
\def\thebibliography{\DeclareRobustCommand{\VAN}[3]{##3}\VANthebibliography}

\usepackage{graphicx}	
\usepackage{amsmath}	

\usepackage{xspace}

\newcommand\sfeff{\ensuremath{\epsilon_{\rm ff}}\xspace}   
\newcommand\Sigmacloud{\ensuremath{\Sigma_{\rm cloud}}\xspace} 
\newcommand\Mcl{\ensuremath{M_{\rm cl}}\xspace} 
\newcommand\taudisc{\ensuremath{\tau_{\rm disc}}\xspace}  
\newcommand{\msun}{\ensuremath{\rm M_{\sun}}\xspace}  

\title[Disc evolution in star clusters]{Star cluster formation from turbulent clumps. IV.  Protoplanetary disc evolution}

\author[Gautam et al.]{
Aayush Gautam,$^{1,2}$\thanks{E-mail: gravitas908@gmail.com}
Juan P. Farias$^{3,4}$
and Jonathan C. Tan$^{2,3}$
\\
$^{1}$ Department of Physics, Birendra Multiple Campus, Tribhuvan University, Bharatpur 44207, Nepal \\
$^{2}$ Department of Astronomy, University of Virginia, Charlottesville, VA 22904, USA\\
$^{3}$ Department of Space, Earth \& Environment, Chalmers University of Technology, Gothenburg SE-41293, Sweden \\
$^{4}$ Department of Astronomy, University of Texas at Austin, TX 78712, USA
}

\date{Accepted XXX. Received YYY; in original form ZZZ}

\pubyear{\the\year{}}

\begin{document}
\label{firstpage}
\pagerange{\pageref{firstpage}--\pageref{lastpage}}
\maketitle

\begin{abstract}
Most stars are born in the crowded environments of gradually forming star clusters. Dynamical interactions between close-passing stars and the evolving UV radiation fields from proximate massive stars are expected to sculpt the protoplanetary discs in these clusters, potentially contributing to the diversity of planetary systems that we observe. Here, we investigate the impact of cluster environment on disc demographics by implementing simple protoplanetary disc evolution models within $N$-body simulations of gradual star cluster formation. We consider a range of star formation efficiency per free-fall time, $\epsilon_{\rm ff}$, and mass surface density of the natal cloud environment, $\Sigma_{\rm cl}$, both of which affect the overall duration of cluster formation. We track the interaction history of all stars to estimate the dynamical truncation of the discs around stars involved in close encounters. We also track external photoevaporation of the discs due to the ionizing radiation field of the nearby high- and intermediate-mass ($> 5 M_\odot$) stars. We find that $\epsilon_{\rm ff}$, $\Sigma_{\rm cl}$, and the degree of primordial binarity have major influences on the masses and radii of the disc population. In particular, external photo-evaporation has a greater impact than dynamical interactions in determining the fate of discs in our clusters. 
\end{abstract}




\section{Introduction}
\label{sec:intro}

Protoplanetary discs (PPDs) are a natural outcome of the process of star formation, forming an important link between early-stage, protostellar accretion discs and planetary systems \citep{Williams2011ARA&A..49...67W,Andrews2020ARA&A..58..483A}. It is widely accepted that most stars are formed together in groups of a wide range of sizes and scales \citep{Lada2003ARA&A..41...57L, Gutermuth2009ApJS..184...18G}. This means the star formation environment has the potential to shape PPDs around these stars from an early stage. Therefore, while the evolution of isolated discs has been extensively studied, it is also important to study PPD evolution within the context of a forming star cluster. 

For this, we have to consider important mechanisms that can affect discs in the dynamic environment of ongoing star formation. First of all, circumstellar discs around stars with close binary partners have been observed to have smaller disc masses and disc radii than those in wide binaries \citep{Offner2022arXiv220310066O, Zurlo2023EPJP..138..411Z}. One way this could occur is if stars are born with binary partners or in a multiple system, these companions can truncate discs around each other during their formation, which we refer to as \emph{primordial binary truncations}.  Similarly, as more stars that will form the future cluster are born, there could be dynamical interactions between them, potentially leading to disc truncations. \cite{Vincke2016ApJ...828...48V} showed simulated discs are smaller in dense clustered environments having more dynamical encounters. We refer to disc truncations during dynamical encounters as \emph{dynamical truncations}. Moreover, if massive stars form in the cluster, they will emit UV photons that ionize and disperse the parent star-forming cloud, exposing the nearby PPDs to the radiation field of the star cluster. The external irradiation will ionize and heat the surface of the discs and drive mass loss through winds, leading to external photoevaporation and systems that appear as ``proplyds'' \citep[e.g.,][]{Johnstone1998ApJ...499..758J}. In this work, we will refer to disc truncations due to external photoevaporation as \emph{radiative truncations}. Finally, various processes local to a host star and its disc are expected to lead to gradual dissipation of the disc, including viscous accretion and internal photoevaporation \citep[e.g.,][]{Gorti2009ApJ...690.1539G}. 

It is important to note that even within the same star cluster, different disc truncation mechanisms may have a dominant influence on disc evolution at different times of the star cluster formation. While the cluster is still embedded in the parent gas cloud, the discs are protected from radiation by surrounding gas whereas close encounters are more common. After expansion and dissolution of the cluster, rates of stellar interactions are reduced, but discs may be more exposed to external radiation due to removal of gas. Furthermore, the initial conditions for the formation of each cluster, such as stellar density, velocity dispersion and number of nearby massive stars could shape the fate of discs in the cluster. The question of which evolutionary processes dominate at different times and environmental conditions has therefore become a field of active research \citep[see, e.g., review by][]{Reiter2022arXiv220903889R}.

In the context of \emph{dynamical truncations}, \cite{Pfalzner2005A&A...437..967P} performed the first detailed parameter study of effects of star-disc encounters on low-mass discs. Investigating the change in disc sizes and mass and angular momentum loss during the encounters, they found that for the strongest impact on discs, the perturber should be massive with respect to the host, and the encounter should be at a small periastron distance, coplanar and parabolic. This is to be expected because a large mass ratio (perturber/host) encounter at a close distance leads to larger angular momentum loss. Similarly, a coplanar parabolic encounter maximizes the interaction time. \cite{Cuello2022arXiv220709752C} reviewed both numerical simulations and observations related to the effect of stellar encounters on planet forming discs. Recently, \cite{Pfalzner2021ApJ...921...90P} performed $N$-body simulations to study the frequency of close flybys in young star clusters. 

Similarly, disc dispersal through radiative truncations in star clusters has been extensively studied. \cite{Armitage2000A&A...362..968A} first studied the influence of clustered star formation environments on disc lifetimes by modeling the effects of extreme UV (EUV) radiation fields due to massive stars, finding that efficient disc dispersal by EUV-driven external photo-evaporation could suppress giant planet formation via the gas accretion route. Similarly, \cite{Qiao2022MNRAS.512.3788Q} investigated the evolution of discs in Carina-like massive star-forming regions focusing on the effects of external photoevaporation. In order to calculate the FUV field strengths responsible for external photo-evaporation of discs, they considered both geometric dilution (inverse-square) of radiation, and Monte Carlo radiative transfer to account for shielding of the radiation by the star-forming gas. Radiation shielding of discs has been proposed as an explanation for the discovery of two distinct disc populations in the NGC 2024 region \citep{vanTerwisga2020A&A...640A..27V}. In this region, discs embedded in the dense natal gas in eastern side are more massive than discs in western side which are readily exposed to the massive star. The impact of such radiation shielding provided by the gas to PPDs in an embedded cluster was also studied recently by \cite{Wilhelm2023arXiv230203721W} who coupled magneto-hydrodynamic simulations of star cluster formation with disc evolutionary models. Recently, \cite{Coleman2022MNRAS.514.2315C} identified different evolutionary pathways for PPDs under both internal photo-evaporation and external photoevaporation. When internal photoevaporation dominates, the discs deplete inside-out whereas discs deplete from outside-in under stronger influence of external photoevaporation. In the parameter space of X-ray luminosity of central star and external photo-evaporative massloss rate, they also found a more complex evolution in the intermediate regime where both mechanisms are equally dominant.

Previous works have attempted in different ways to include the effects of some (or all) of these processes on circumstellar or protoplanetary discs in star clusters. With a statistical Monte-Carlo approach, \cite{Winter2018MNRAS.478.2700W} studied the role of stellar densities in determining which of external photo-evaporation and dynamical encounters would be more dominant in a star cluster. \cite{Concha2023MNRAS.520.6159C} simulated the evolution of circumstellar discs in young star clusters in the presence of external photo-evaporation and dynamical encounters. Similarly, \cite{Marchington2022MNRAS.515.5449M} studied the evolution of disc masses and disc radii in star forming regions with a post-processing analysis of $N$-body simulations, including external photo-evaporation and viscous evolution.

However, it remains a challenge to model all of these processes in full detail in an on-the-fly simulation of ongoing formation of a star cluster while fully resolving stellar dynamics and binaries. Most of the previous works either assume that the stars are formed at once, or have to post-process star cluster formation simulations. In our approach, we follow the evolution of discs as their host stars are formed and orbit in an ongoing $N$-body star cluster formation simulation, where 50\% stars are formed in primordial binaries. We couple semi-analytical models of disc evolution to each star in the simulation and track the evolution of disc mass and disc radii. 

The paper is structured as follows: we review our models of gradual star cluster formation in section \ref{subsec:starclusterformation}. Then, we discuss the disc models in section \ref{subsec:discmodel} followed by the implementation of dynamical truncations and external photo-evaporation in sections \ref{subsec:dynamicalfeedback-primordialfeedback} and \ref{subsec:radiativefeedback}, respectively. In the results section \ref{sec:results}, we present how disc masses and radii are affected by formation timescales of the cluster and compare our simulated disc demographics with observations of discs in the Orion and G286 star forming regions. We discuss the main findings and caveats of our modeling in the discussion section \ref{sec:discussion}. 

\section{Methods}
\label{sec:methods}

In this paper, we aim to study the effect of different mechanisms that can truncate protoplanetary discs (PPDs) during the star cluster formation process. We have implemented simple semi-analytical models of PPD evolution inside $N$-body models of gradual star cluster formation. Disc models evolve together with the trajectory of each star, allowing us to resolve each stellar encounter with full time resolution. We describe both the star cluster formation simulations and coupled PPD models in the following sub-sections.

\subsection{Star Cluster Formation Simulations} 
\label{subsec:starclusterformation}

We simulated PPD evolution within star cluster formation models developed by \citet{Farias2017ApJ...838..116F}, \citet{Farias2019MNRAS.483.4999F} and \citet{Farias2023arXiv230108997F}. These consist of a suite of $N$-body simulations focused on accurately following stellar dynamics in the early stages of star cluster formation, i.e., where there is still ongoing formation of stars. These models are based on the Turbulent Clump Model \citep[TCM;][]{McKee2003ApJ...585..850M}, where star clusters are formed inside magnetized, turbulent, gravitationally bound, and initially starless clumps within giant molecular clouds (GMCs). \cite{Farias2017ApJ...838..116F} (hereafter, Paper I) presented the extreme case of instantaneous star cluster formation, where all stars are formed together at once and remaining gas is assumed to leave the system instantaneously according to a given global star formation efficiency ($\epsilon$). \cite{Farias2019MNRAS.483.4999F} (hereafter, Paper II) presented a more realistic case of star cluster formation, where stars are formed gradually, parameterized by the star formation efficiency per free-fall time (\sfeff), and a corresponding amount of natal gas is expelled gradually. We implemented our models of PPD evolution within these $N$-body models outlined in Paper II. 

In these pure $N$-body models, we emulate the parent cloud gas using an analytical potential that follows the structure from the TCM model. The star cluster is formed inside the turbulent gas clump of mass \Mcl embedded in the surrounding molecular cloud with mass surface density \Sigmacloud. The clump is modeled as a polytropic sphere of gas with a density profile of the form:
\begin{eqnarray}
    \rho_{\rm cl} (r) = \rho_{\rm s,cl} \left( \frac{r}{R_{\rm cl}} \right)^{-k_{\rho}},
    \label{eqn:polytropicdensity}
\end{eqnarray}
where $\rho_{\rm s,cl}$ and $R_{\rm cl}$ are the density at the surface of the clump and the radius of the clump, respectively. Similarly, the velocity dispersion profile of the clump is given by:
\begin{eqnarray}
    \sigma_{\rm cl} (r) = \sigma_{\rm s} \left( \frac{r}{R_{\rm cl}} \right)^{(2-k_{\rho})/2},
    \label{eqn:polytropicvelocitydispersion}
\end{eqnarray}
where $\sigma_{\rm s}$ is the velocity dispersion at the surface of the clump. We set the exponent $k_{\rho}$ = 1.5 following Paper II. The time taken for the star cluster to form inside the clump is given by:
\begin{eqnarray}
    t_\star = \frac{\epsilon}{\sfeff} t_{\rm ff, 0} \quad,
    \label{eqn:formationtimescale}
\end{eqnarray}
where $t_{\rm ff,0}$ is the initial global freefall timescale of the clump given by:
\begin{eqnarray}
    t_{\rm ff,0} = 0.069 \left( \frac{\Mcl}{3000 M_\odot}\right)^{1/4} \left( \frac{\Sigmacloud}{\rm g\: cm^{-2}}\right)^{-3/4} \rm{Myr} \quad,
    \label{eqn:freefalltimescale}
\end{eqnarray}
in the TCM model (Equation 9, Paper II). From equation \ref{eqn:formationtimescale}, we see that the star cluster formation time ($t_\star$) is shorter for increasing values of \sfeff with the limiting case of instantaneous formation represented by $\sfeff=\infty$. Moreover, for the same value of $\epsilon_{\rm ff}$, $t_\star$ is shorter for higher \Sigmacloud, because the freefall timescale $t_{\rm ff,0}$ is shorter in higher \Sigmacloud environments. As stars are gradually introduced, the background gas mass dissipates linearly matching the global $\epsilon$ and \sfeff.

For our simulations, we fix \Mcl to $3000\:\msun$ and $\epsilon$ to 0.5. We consider star clusters formed from clumps in two different environments: M3000L set formed from clumps with low surface density ($\Sigma_{\rm cloud}= 0.1\: \rm g\: cm^{-2}$), and M3000H set formed from clumps with high surface density ($\Sigma_{\rm cloud}= 1.0\: \rm g\: cm^{-2}$). For each \Sigmacloud, we consider six values for $\epsilon_{\rm ff}$: 0.01, 0.03, 0.10, 0.30, 1.00 and $\infty$. Table \ref{tab:simulations} shows the full set of cluster parameters (\sfeff and \Sigmacloud) modeled in this work. For each row of Table \ref{tab:simulations} we have run 5 realizations, each realization with an IMF sampling \citep{Kroupa2001MNRAS.322..231K} and 50\% primordial binary population as described in \cite{Farias2023arXiv230108997F}.

\begin{table*}
\centering
\caption{Overview of star cluster formation simulations}
\label{tab:simulations}
\begin{tabular}{rcccccccc}
Set Name & \sfeff & \Sigmacloud & \Mcl & $\langle N_* \rangle$ & $t_{\star}$ & $t_{\rm ff}$ & $R_{\rm cl}$ & $\sigma_{s}$\\ 
& & [$\rm g\:cm^{-2}$] & [$M_\odot$]  &   & [Myr] & [Myr] & [pc] & [$\rm km\: s^{-1}$] \\ \hline

  & 0.01 & 0.1 & 3,000  & 4,000  & 19.40 & 0.39  & 1.15 & 1.71 \\
  & 0.03 & 0.1 & 3,000  & 4,000  & 6.47  & 0.39  & 1.15 & 1.71 \\
M3000L  & 0.1  & 0.1 & 3,000  & 4,000  & 1.94  & 0.39  & 1.15 & 1.71 \\
  & 0.3  & 0.1 & 3,000  & 4,000  & 0.65  & 0.39  & 1.15 & 1.71 \\
  & 1.0  & 0.1 & 3,000  & 4,000  & 0.19  & 0.39  & 1.15 & 1.71 \\
  & $\infty$ & 0.1 & 3,000 & 4,000 & 0.00 &  0.39 & 1.15 & 1.71 \\ \cline{1-9}
 &  0.01 & 1.0 & 3,000  & 4,000  & 3.45  & 0.069  & 0.365 & 3.04 \\
 &  0.03 & 1.0 & 3,000  & 4,000  & 1.15  & 0.069  & 0.365 & 3.04 \\
M3000H &  0.1  & 1.0 & 3,000  & 4,000  & 0.35  & 0.069  & 0.365 & 3.04 \\
 &  0.3  & 1.0 & 3,000  & 4,000  & 0.12  & 0.069  & 0.365 & 3.04 \\
 &  1.0  & 1.0 & 3,000  & 4,000  & 0.03  & 0.069  & 0.365 & 3.04 \\
 & $\infty$ & 1.0 & 3,000 & 4,000 & 0.00 & 0.069 & 0.365 & 3.04 \\ \cline{1-9}
\end{tabular}
\end{table*}

\subsection{Disc Model} \label{subsec:discmodel}
We assign a PPD to every star in the star cluster at its formation, with the disc mass being 10\% of host star mass. We assign an individual disc to each member of a primordial binary - however, we do not consider circumbinary discs. We define current age of an individual disc as $t_{\rm{age}} = t - t_{\rm{f}}$, where $t$ is the global simulation time and $t_{\rm{f}}$ is the time of formation of the host star. Then, the disc mass when $t_{\rm{age}} = 0$ is given by:
\begin{eqnarray}
    m_{\rm{disc}} = 0.1 m_{\rm{star}},
    \label{eqn:initialdiscmass}
\end{eqnarray}
where $m_{\rm{star}}$ is the mass of the host star at its formation. We assume that the mass of each disc is distributed radially following a power-law surface density profile of the form:
\begin{eqnarray}
    \Sigma (r) = \Sigma_0 \left(\frac{r}{\rm r_0}\right)^{-p} &; r_{\rm in} \leq r \leq r_{\rm out},
    \label{eqn:initialsurfacedensity}
\end{eqnarray}
truncated at an inner radius $r_{\rm in} = 0.01$ AU and outer radius $r_{\rm out}= 100$ AU. We use the same $r_{\rm in}$ and $r_{\rm out}$ for all discs, and set the scale radius $r_0$ to 1 AU. The value of the constant exponent $p$ is rather uncertain in observations while previous theoretical works also assume different values \citep[see, e.g.,][]{Steinhausen2012A&A...538A..10S} ranging from $p$ = 0.75 to 1.5. For our disc models, we have to opted to use the value $p = 0.8$.
Finally, $\Sigma_0$ is a normalization constant set at formation by the initial disc mass and for $p \neq 2$, it is given by: 
\begin{equation}
    \Sigma_0 = \frac{ (2-p) m_{\rm{disc}}}{2\pi r_0^2}  \left[ \left(\frac{r_{\rm out}}{r_0}\right)^{2-p} \left(1- \left(\frac{r_{\rm in}}{r_{\rm out}}\right)^{2-p}\right) \right]^{-1}.
    \label{eqn:initialsurfacedensityconstant}
\end{equation}

We evolve discs using an exponential decay of mass with a characteristic lifetime \taudisc = 2 Myr, while keeping their radius unchanged. The surface density at disc age $t_{\rm{age}}$ is given by:
\begin{equation}
    \Sigma (r,t_{\rm{age}})= \Sigma_0 \left( \frac{r}{r_0}\right)^{-p}  e^{-\frac{t_{\rm{age}}}{\taudisc}}.
    \label{eqn:evolvingsurfacedensity}
\end{equation}
Finally, the current mass of the disc having outer radius $r_{\rm out}$ at $t_{\rm{age}}$ is given by:
\begin{eqnarray}
    m_{\rm{disc}} (r_{\rm out},t_{\rm{age}}) &=& \frac{2\pi}{2-p} \Sigma (r, t_{\rm{age}}) \left( \frac{r_{\rm out}}{r_0} \right)^{2-p} \nonumber \\ 
    & & \times  \left( 1-\left(\frac{r_{\rm in}}{r_{\rm out}} \right)^{2-p} \right).
    \label{eqn:currentmass}
\end{eqnarray}

\subsection{Dynamical truncations}
\label{subsec:dynamicalfeedback-primordialfeedback}

We model the effect of stellar encounters as an additional source of mass loss via dynamical truncations of the discs. Stars will perturb discs during stellar encounters leading to dynamical truncations of discs. In general, considering the encounter of the disc bearing star of mass $m_{\rm host}$ with a perturbing star $m_{\rm pert}$ at a periastron distance $d_{\rm peri}$, a disc particle around the host star could be assumed to be stable up to the Hill radius ($r_{\rm hill}$) of the host star \citep{Hamilton1991Icar...92..118H} given by:
\begin{equation}
  r_{\rm hill} = d_{\rm peri} \left( \frac{m_{\rm host}}{3m_{\rm pert}}\right)^{1/3}.
  \label{eqn:hillradius}
\end{equation}

However, \cite{Breslau2014A&A...565A.130B} studied the effect of stellar encounters on the sizes of PPDs by simulating prograde, coplanar, and parabolic encounters over a range of mass ratios and periastron distances observed in typical clusters. They obtained that the resulting truncation radius can be expressed as a function of the mass ratio and periastron distance between the host star and perturbing star as:
\begin{equation}
    r_{\rm new} = 0.28 \times d_{\rm peri} \left( \frac{m_{\rm host}}{m_{\rm pert}}\right)^{0.32},
    \label{eqn:breslau}
\end{equation}
which is approximately 40\% of the Hill radius calculated for the same encounter from equation \ref{eqn:hillradius}. Equation \ref{eqn:breslau} is averaged over all inclinations of stellar encounters, so we do not consider the orientation of the discs during the stellar encounters. In short, we use the \cite{Breslau2014A&A...565A.130B} criterion for calculating dynamical truncations of discs in our simulations. 

For the initial unperturbed disc, the disc size is the outer disc radius $r_{\rm out}$. However, after each stellar encounter, we update $r_{\rm out}$ to the new truncation radius $r_{\rm new}$ if $r_{\rm new} < r_{\rm out}$. In this case, we remove the disc material beyond $r_{\rm new}$ entirely from the simulation - we do not consider accretion of the lost disc material onto either the host or perturbing star. 

Finally, in addition to dynamical truncations during stellar encounters, discs formed around stars in binaries could be subject to perturbations from binary partners during and after formation of disc, leading to truncations. We refer to this contribution to mass-loss as primordial binary truncations. For this, we calculate the truncation radii given by equation \ref{eqn:breslau}, assuming the binary partner as the perturber. Therefore, both stars perturb discs around their partner, potentially leading to truncation to smaller disc radii and disc masses at formation itself.   

\subsection{Radiative truncations}
\label{subsec:radiativefeedback}

We model the external photoevaporation of PPDs due to ionizing radiation fields in the star cluster. At the location of each disc, we calculate the contribution of each massive star ($m_{\star} \geq 5 M_\odot$) to the EUV radiation field. These EUV radiation fields drive photoevaporative mass loss in the outer regions of the disc. However, discs may be shielded from the effects of EUV radiation by the background gas present during the formation of the cluster. Therefore, we model the shielding of discs by considering Str\"omgren spheres around each of the massive stars. A Str\"omgren sphere is a steady-state idealization of an fully ionized HII region around a hot, luminous star, where the photoionization is balanced by hydrogen recombination \citep[e.g.,][]{draine2010physics}. We assume that the disc will be subject to the ionizing radiation of the massive star only if it is within the Str\"omgren sphere of the star. The radius of Str\"omgren sphere around a massive star is given by:
\begin{eqnarray}
    R_S &=& \left( \frac{3}{4\pi} \frac{\Phi_\star}{(n_{\rm H})^2 \alpha_B} \right)^{1/3} \nonumber \\
        &=& 9.77 \times 10^{18} (\Phi_{49})^{1/3} \times n_2^{-2/3} \times T_4^{0.28} \:{\rm cm},
    \label{eqn:Stromgrenradius}
\end{eqnarray}
where $\Phi_{49} = \Phi_\star/10^{49} {\rm s}^{-1}$, $T_4 = T/10^4 {\rm K}$, $n_2=n_{\rm H}/10^2  {\rm cm}^{-3}$ and $\alpha_B$ is the thermal rate coefficient for hydrogen recombination. We obtain EUV luminosities $\Phi_\star$ as a function of stellar mass from \cite{Armitage2000A&A...362..968A}. We calculate the hydrogen number density ($n_{\rm H}$) from the gas density within the half-mass radius of the cluster. In addition, we assume that the ionizing radiation is attenuated with distance $d$ from the massive star by a factor $\left( 1-\frac{d^3}{R_S^3} \right)$ inside the HII region. 

Then, for a disc of radius $r_{\rm disc}$ at a distance $d < R_S$ from a massive star, the photoevaporative mass-loss is given by \citep{Hollenbach2000prpl.conf..401H}:
\begin{eqnarray}
    \dot{m}_{\rm EUV} &=& 7.0 \times 10^{-9} \left( \frac{ \left( 1-\frac{d^3}{R_s^3} \right) \Phi_\star}{10^{49} {\rm s}^{-1}} \right)^{1/2} 
    \left( \frac{r_{\rm disc}}{10 {\rm AU}} \right)^{3/2} \nonumber \\  & & \times \left(\frac{d}{10^{17} {\rm cm}} \right)^{-1} M_{\odot} {\rm yr}^{-1}. 
    \label{eqn:evaporationmasslossrate}
\end{eqnarray}
We sum up the mass loss due to every massive star. We implement this mass loss due to external photo-evaporation as a truncation of the disc up to the radius that encloses the new mass obtained by subtracting the calculated massloss during the timestep. We note that there is a critical gravitational radius ($r_{g}$), where the sound speed of the ionized gas ($c_s$) is equal to the orbital velocity. The dust and gas at $r < r_{g}$ are bound to the host star. The photoevaporative mass loss occurs only at radii $r > r_{g}$. The gravitational radius for a disc around host star of mass ($m_\star$) is given by \citep{armitage_astrophysics_2020}:
\begin{equation}
    r_{g} = \frac{Gm_\star}{c^2_{s}} \approx 8.9 \left(\frac{m_\star}{M_\odot}\right) \left(\frac{c_{s}}{10 {\rm km s}^{-1}}\right)^{-2} {\rm AU},
    \label{eqn:gravitationalradius}
\end{equation}
where $c_{s} = 10\: {\rm km\: s}^{-1}$. The disc evaporates from outside-in up to $r_{\rm g}$. Once the disc becomes smaller than the radius $r_{\rm g}$, external photo-evaporation gets stopped for this particular disc.

\subsection{Implementation}
\label{subsec:implementation}
In our fiducial simulation, we introduce discs at the time of formation of their host stars: we assign each disc a mass equal to 10\% of host star mass and a size of 100 AU. If the host star has a binary partner, we check if the disc should be truncated. If the criterion is satisfied, we truncate the disc to a smaller disc mass and disc radius before the next time-step. After their formation, the discs undergo an exponential decay ($\taudisc=2\:$Myr) every time step. At every time step, we also check for possible dynamical truncations of the disc by nearby stars. In addition, we calculate the expected mass-loss due to the radiation at the location of the disc. We find the new disc radius taking both dynamical and radiative truncations into account. Then, we update the mass and radius of the discs before starting the next time step. We implemented the above treatment of discs within a modified version of Nbody6++ \citep{aarseth2003gravitational,Wang2015MNRAS.450.4070W} developed to model gradual star cluster formation (Nbody6SF, Paper II).

\section{Results}
\label{sec:results}

\subsection{Evolution of the mean disc properties}
\label{subsec:meandiskproperties}

\begin{figure*}
    \includegraphics[width=\textwidth]{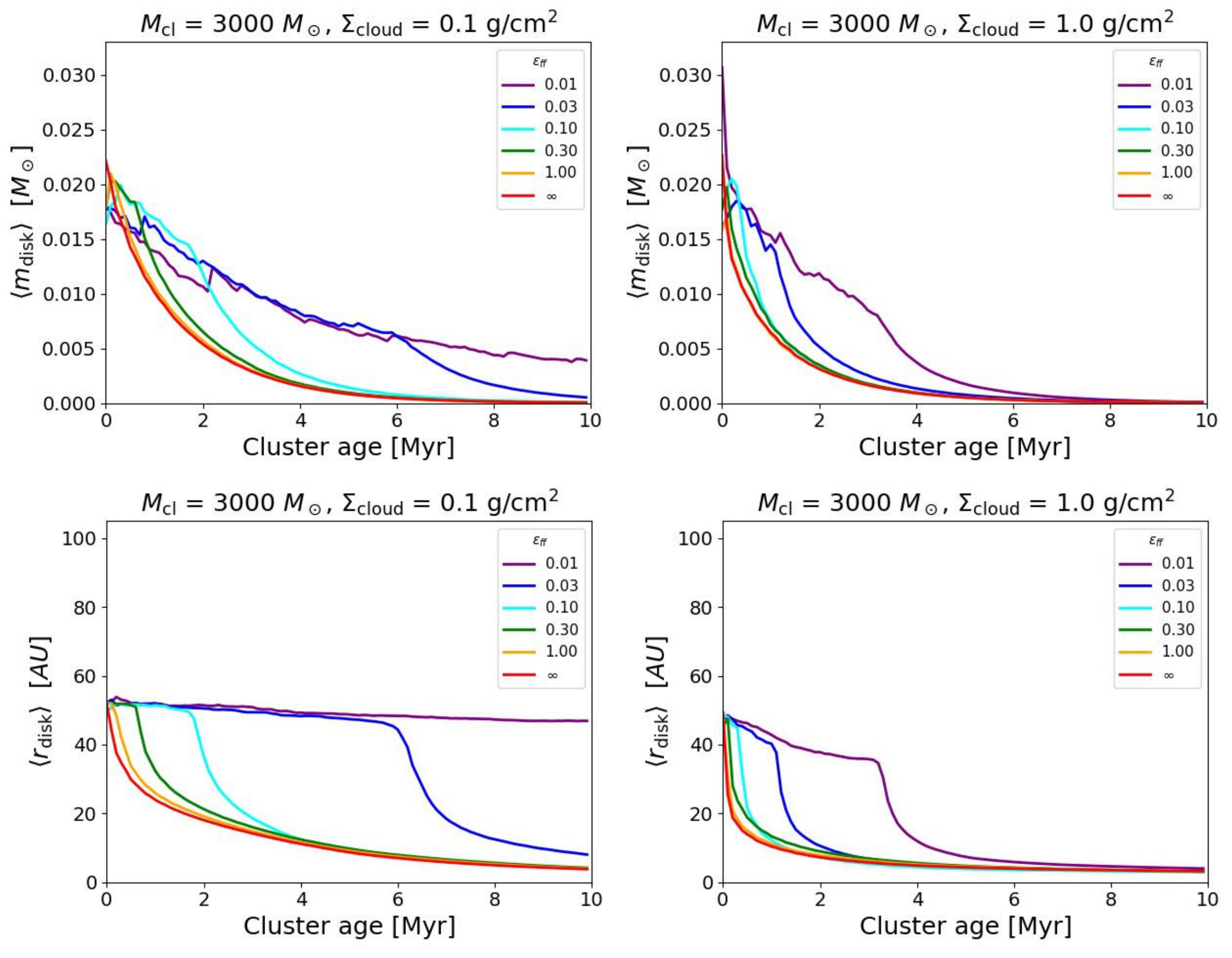}
    \caption{The evolution of mean disc mass, $\langle m_{\rm disk} \rangle$, (top row) and mean disc radius, $\langle r_{\rm disk} \rangle$, (bottom row) for the simulated clusters formed from clumps inside low density parent clouds ($\Sigma_{\rm cloud}=0.1\:\rm g\:cm^{-2}$) (left column) and high density parent clouds ($\Sigma_{\rm cloud}=1.0\:\rm g\:cm^{-2}$) (right column). The different \sfeff models are represented by the color scheme (purple: 0.01, blue: 0.03, cyan: 0.10, green: 0.30, orange: 1.00 and red: $\infty$). This color scheme is followed consistently throughout this work.}
\label{fig:meandiskpropertyevolution_clusterage}
\end{figure*}

\begin{figure*}
    \centering
    \includegraphics[width=\textwidth]{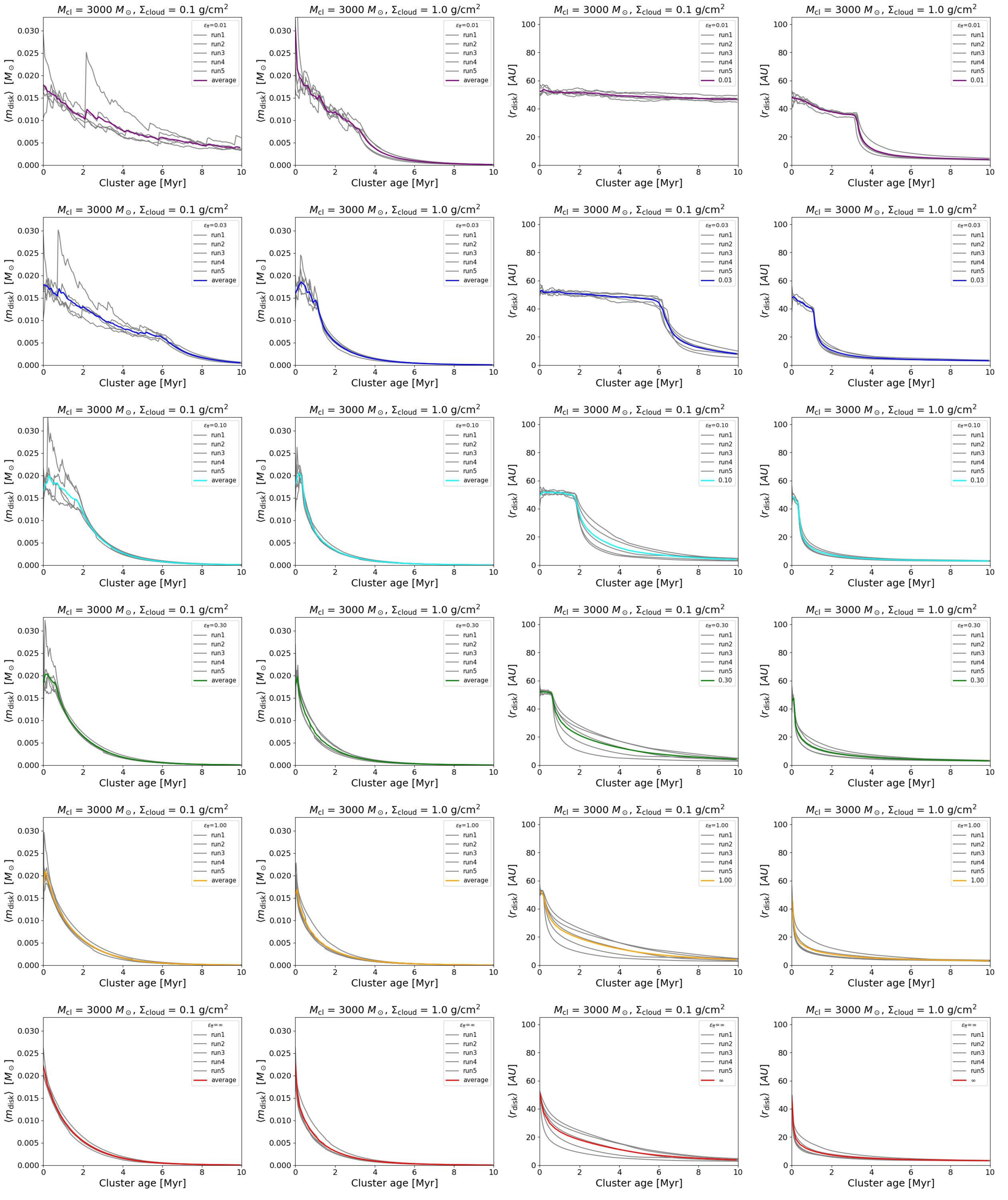}
    \caption{The evolution of mean disc mass and mean disc radius for the full set of models discussed in section \ref{subsec:meandiskproperties}. The grey lines show the individual runs for each model and the solid colored lines show the average of the mean disc masses (columns 1-2) and mean disc radius (columns 3-4) calculated from the runs.}
    \label{fig:meandiskpropertiesallmodels}
\end{figure*}

The evolution of mean disc mass, $\langle m_{\rm disk} \rangle$, and mean disc radius, $\langle r_{\rm disk} \rangle$, in the simulated clusters over a period of 10 Myr are shown in the top and bottom rows, respectively, in Figure~\ref{fig:meandiskpropertyevolution_clusterage}. The left column shows the case of clusters formed from clumps in the low density cloud environment (set M3000L), while the right column shows the high density cloud environment case (set M3000H). The full set of models is shown in Figure \ref{fig:meandiskpropertiesallmodels}

We see from the bottom row of Fig.~\ref{fig:meandiskpropertyevolution_clusterage}, that the primordial binaries strongly influence the initial mean disc radius. We have 50\% primordial binaries in the simulations. For discs present around stars in binaries, the gravitational influence of the binary partner truncates the disc at the radius given by equation \ref{eqn:breslau}.  Therefore, the mean disc radius starts around 55 AU, instead of the unperturbed size of 100 AU for single discs.


From the top row of Fig.~\ref{fig:meandiskpropertyevolution_clusterage}, we observe that the mean disc mass starts at around $0.02 \: M_\odot$ at $t$ = 0 Myr. While the star formation process is still ongoing, the mean disc mass fluctuates due to formation of new discs in the cluster, but it gradually decreases with time in all the models. The remaining mean disc mass at $t$ = 10 Myr for both M3000L and M3000H sets is around $10^{-6} M_\odot$. 
Starting from the time of formation of the discs, exponential decay, dynamical truncations and external photoevaporation contribute to this disc mass loss. We examine the relative contribution of these mechanisms in section \ref{subsec:fractional_contributions}.

We notice that there is a significant drop in mean disc mass and mean disc radius once star formation is complete in the cluster. The mean disc mass and mean disc radii drop much faster from this moment because once all stars are formed, the background gas is dissipated away. This turns off the shielding of discs from the radiation fields in the cluster and leads to efficient external photo-evaporation.  

\begin{figure*}
    \centering
    \includegraphics[width=\textwidth]{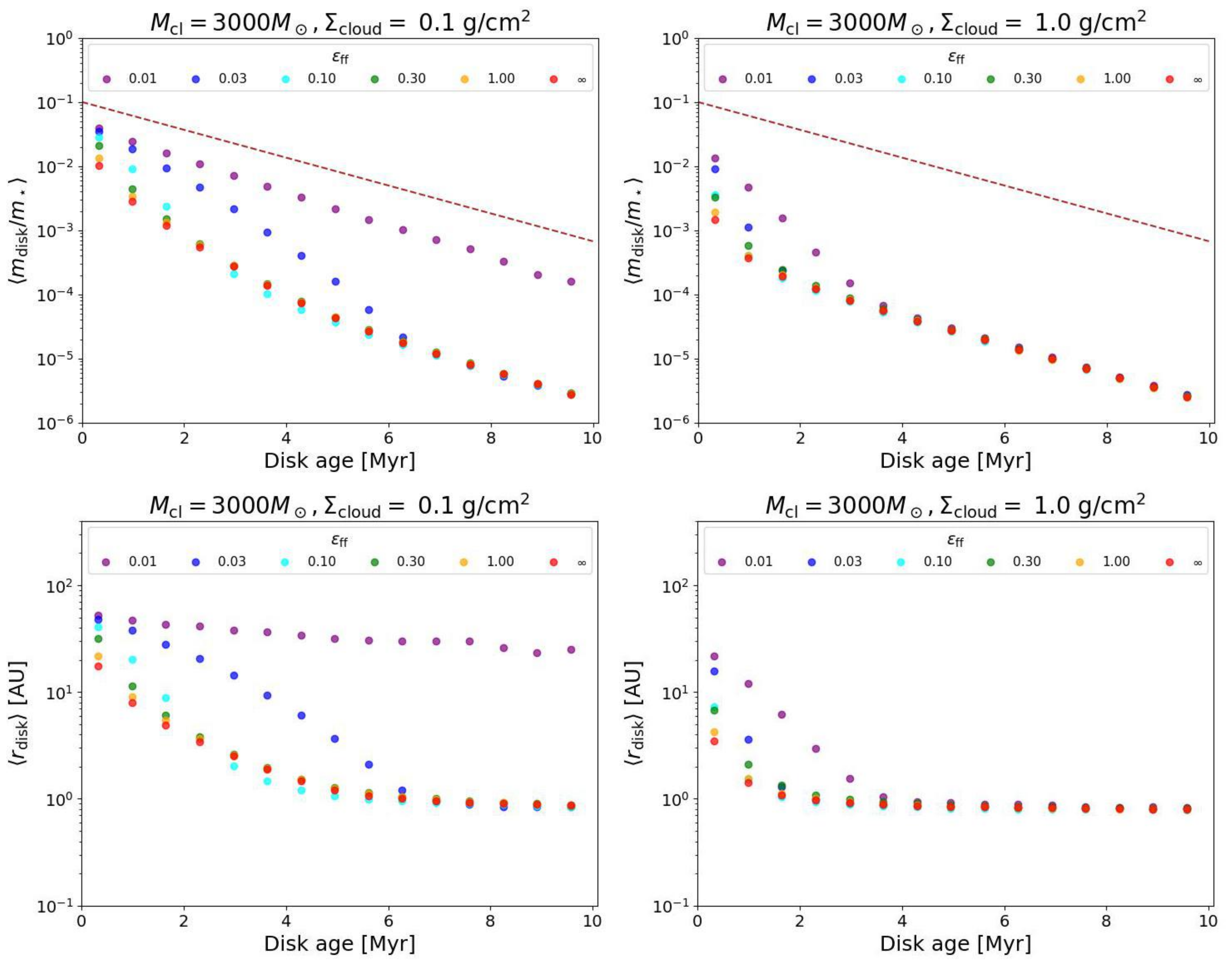}
    \caption{The evolution of mean disc mass normalized by host mass ($\langle m_{\rm disk}/ m_{\rm\star} \rangle$; top row) and mean disc radius ($\langle r_{\rm disk} \rangle$; bottom row) as a function of disc age (in Myr) for the simulated clusters formed from clumps inside low density parent cloud ($\Sigma_{\rm cloud}=0.1\:\rm g\:cm^{-2}$) and high density parent cloud ($\Sigma_{\rm cloud}=1.0\:\rm g\:cm^{-2}$) in the left and right columns, respectively. In the upper panels, the brown dashed line shows the evolution of an isolated disc evolving solely under exponential decay.}
    \label{fig:meandiskpropertyevolution_diskage}
\end{figure*}

We also study the evolution of mean disc properties as a function of the individual disc age, $t_{\rm disk}$. Figure \ref{fig:meandiskpropertyevolution_diskage} shows the mean disc mass (top row) and mean disc radius (bottom row) as a function of the disc age. The dashed line shows the evolution of disc mass under exponential decay for an isolated disk.
As expected, disc mass drops faster under the influence of cluster environment compared to the exponential decay line. We observe that there is a trend of decreasing disc mass and disc radius at the same disc age, when we move to higher \sfeff models. Similarly, at the same disc age, the disc mass and disc radii are smaller for the M3000H set compared to the M3000L set. These trends are expected since higher \sfeff and higher \Sigmacloud generally lead to higher number densities of stars at early times, which could lead to more dynamical truncations and photo-evaporation. We observe that discs at older ages have similar masses and radii for all \sfeff models. However, when they are young ($< 3$ Myr), significant differences can be caused by the varying star formation environment across the \sfeff models. If planet formation happens within this early phase, differences in mass and radius of these discs may lead to variations in properties of planetary systems across the different \sfeff models.  

\subsection{Mass and Radius Distributions}
\label{subsec:PDFdiskproperties}

\begin{figure*}
    \centering
    \includegraphics[width=0.9\textwidth, height=0.9\textheight]{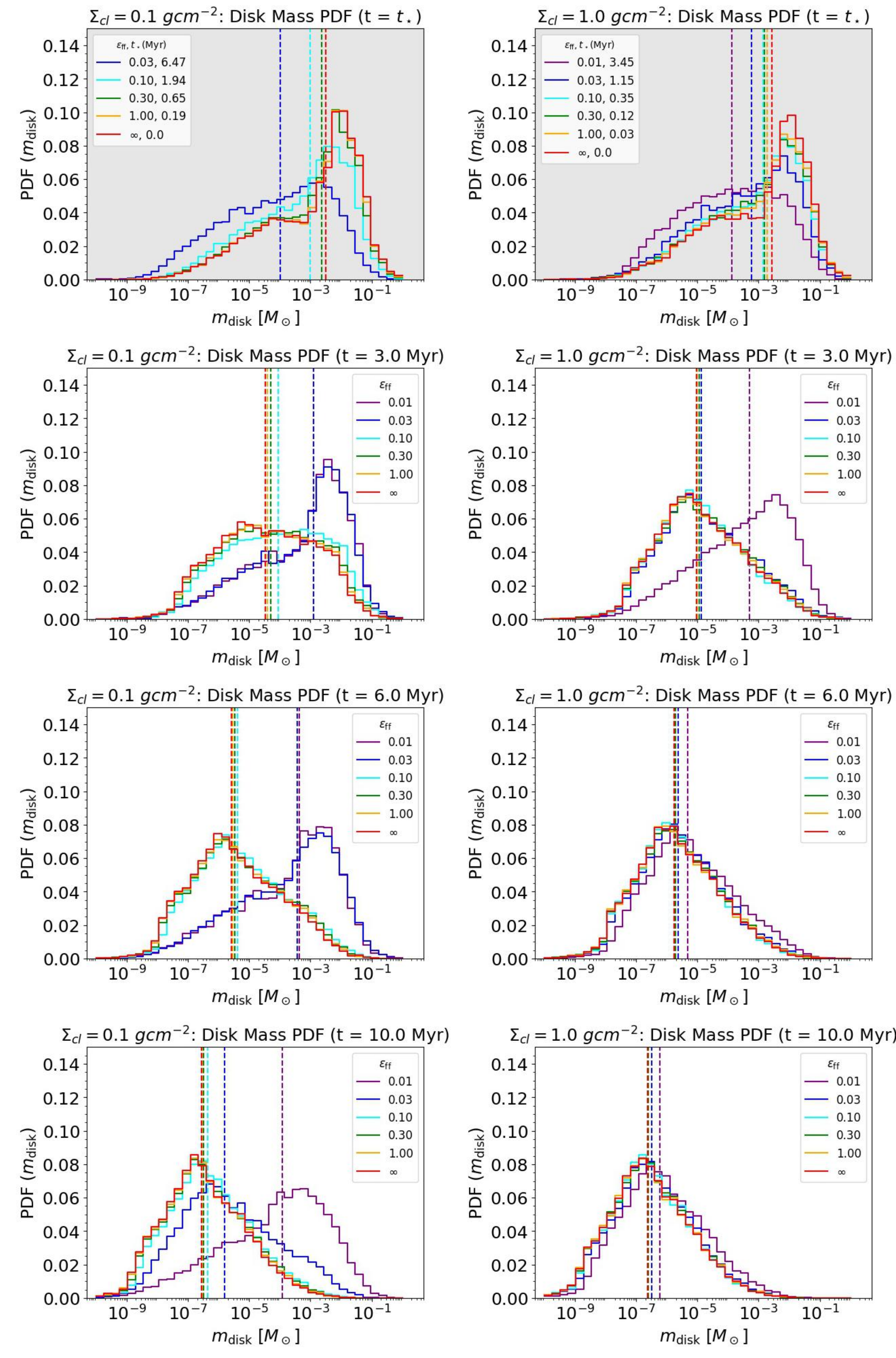}
    \caption{The probability density function (PDF) of the disc masses ($m_{\rm disk}$) for the simulated clusters formed from clumps in low density clouds ($\Sigmacloud =0.1\:\rm g\:cm^{-2}$) and high density clouds ($\Sigmacloud=1.0\:\rm g\: cm^{-2}$). From top to bottom, the PDFs are shown for times $t = t_{\star} =$ 3.0, 6.0 and 10.0 Myr. The dashed vertical lines show the median disc mass for corresponding \sfeff models.}
    \label{fig:diskmasspdf}
\end{figure*}

\begin{figure*}
    \centering
    \includegraphics[width=0.9\textwidth, height=0.9\textheight]{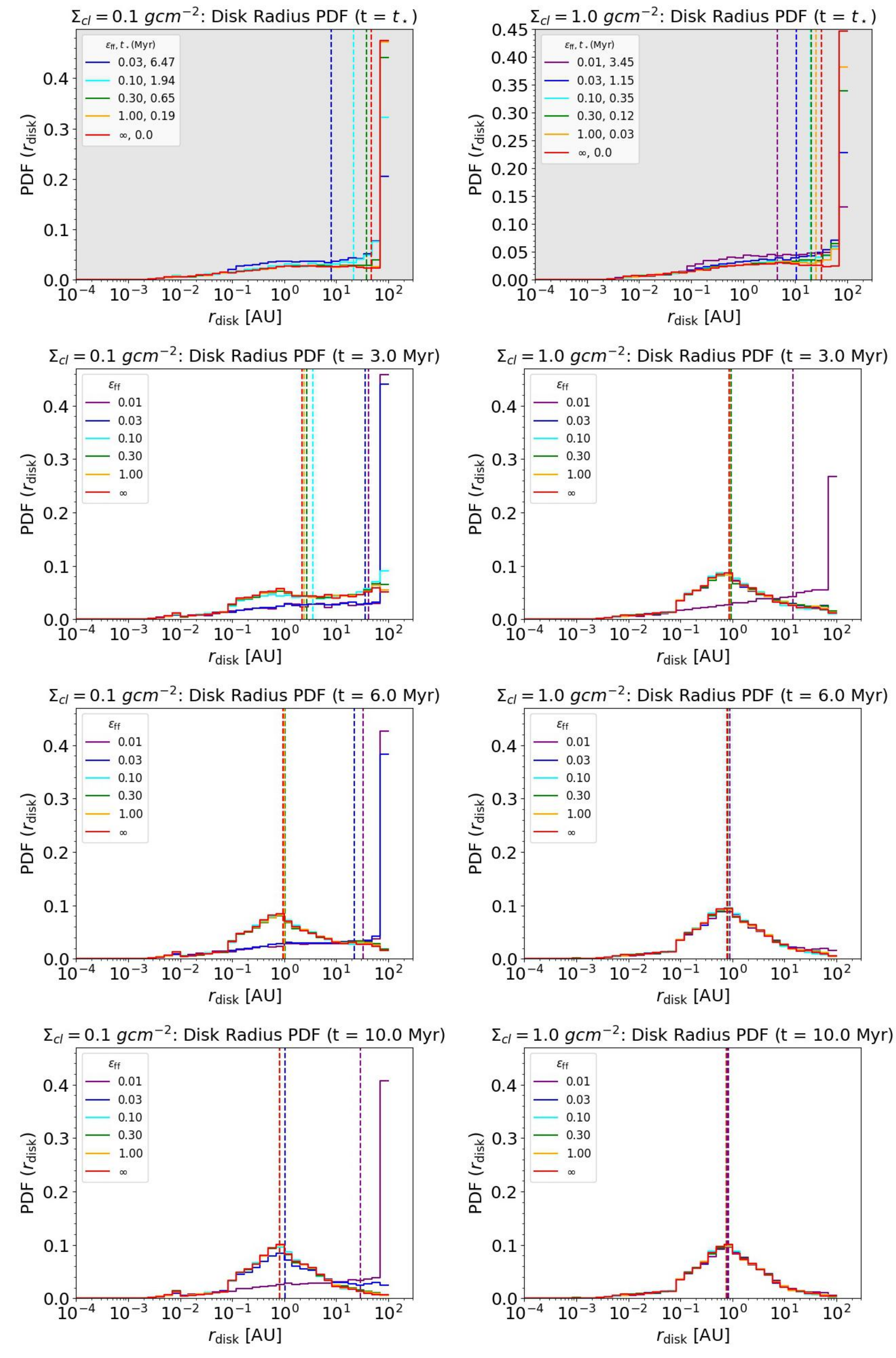}
    \caption{PDF of the disc radii ($r_{\rm disk}$) for the simulated clusters formed from clumps in low density clouds ($\Sigmacloud = 0.1\: \rm g\:cm^{-2}$) and high density clouds ($\Sigmacloud = 1.0\:\rm g\: cm^{-2}$). From top to bottom, the PDFs are shown for times $t = t_{\star}=$ 3.0, 6.0 and 10.0 Myr. The dashed vertical lines show the median disc radius for corresponding \sfeff models.}
    \label{fig:diskradiuspdf}
\end{figure*}

Figure \ref{fig:diskmasspdf} shows the probability distribution functions (PDFs) for disc mass for the different $\epsilon_{\rm ff}$ models at different times $t = t_{\star}$ of 3.0, 6.0 and 10.0 Myr from the start of formation of cluster. Here, $t_{\star}$ is the time when star formation is complete in the particular model. The vertical dashed lines show the median disc mass for the corresponding \sfeff models. The left column shows the distributions for the M3000L clusters, while the right column shows the M3000H clusters. Similarly, Figure \ref{fig:diskradiuspdf} shows the PDF for the disc radii in the same set of simulations.

The top row (shaded) shows the PDF of disc masses at the time when star formation is complete in the cluster ($t=t_\star$). The star formation times for the respective models are mentioned in the legend. We have assigned disc masses equal to 10\% of the host stellar mass. Therefore, the disc initial mass function (and thus, the initial mass PDF) around single stars will have the same shape as the stellar initial mass function, but shifted to smaller masses by a factor of 10. The instantaneous case, shown by the red lines, 
represents this case. In the instantaneous case ($\sfeff = \infty$), all stars are formed at the beginning and so we have the complete set of stars, including the binaries. The effect of binaries in truncating the discs is reflected in the PDF of disc mass. In the other models (\sfeff), the stars are gradually formed and thus the discs are formed at different times. We have the complete set of discs only at $t = t_\star$. This means that if the timescale for star cluster formation is longer, there is more time for feedback mechanisms to act upon the discs. Therefore, the disc masses are lower in general for progressively smaller \sfeff, in which the star clusters take longer times to form.  Accordingly, we observe that for both low and high mass surface densities (\Sigmacloud), the lower \sfeff cases have smaller masses. The case of \sfeff = 0.01 is not shown for the M3000L set because star formation is not complete by 10 Myr for this particular model.  

Now, if we consider PDFs at different cluster ages ($t$ = 3, 6 and 10 Myr), we observe that disc mass PDFs shift to lower masses with increasing cluster age for all \sfeff models due to combined effects of exponential decay, dynamical truncations and external photo-evaporation. Across \sfeff models, the slower forming models (\sfeff = 0.01 and 0.03) show differences from the faster forming models at $t$ = 3.0 and 6.0 Myr in the M3000L set. This difference is attributed to the fact that these models are still forming stars at this point and thus embedded in remnant gas, which shields discs from external photo-evaporation. This is supported by the fact that once star formation is complete after 6 Myr for \sfeff = 0.03 model, its discs experience more photo-evaporation and start to resemble discs in the faster forming models ($\sfeff \geq 0.10$) by $t=10\:$Myr. In contrast, the \sfeff = 0.01 model is still forming stars even at 10~Myr, and its discs remain more massive. In M3000H clusters, star formation is complete in all \sfeff models by 3.5 Myr (see Table \ref{tab:simulations}), so PDFs are roughly similar for different \sfeff models. 

Similarly, if we consider the PDFs of disc radius at the same times as above, we observe that discs in slower-forming \sfeff models are larger than discs in faster-forming \sfeff models while there is ongoing star formation. As the clusters age, disc radii in all \sfeff models approach the $\sfeff = \infty$ distribution. During early phases of cluster evolution (ongoing star formation), we also observe discs of size 100 AU which are not seemingly affected by feedback. However, these discs are still undergoing exponential decay in mass. In our simulations, disc radius is changed only by dynamical truncations and external photoevaporation. So, in an environment shielded from photo-evaporation and having few dynamical truncations, discs could remain large enough for planets to form in the outer regions. However, once star formation is complete, this population of discs is quickly depleted by external photo-evaporation in both M3000H and M3000L clusters.  

\subsection{Fractional Contributions}
\label{subsec:fractional_contributions}

\begin{figure*}
    \centering
    \includegraphics[width=\textwidth]{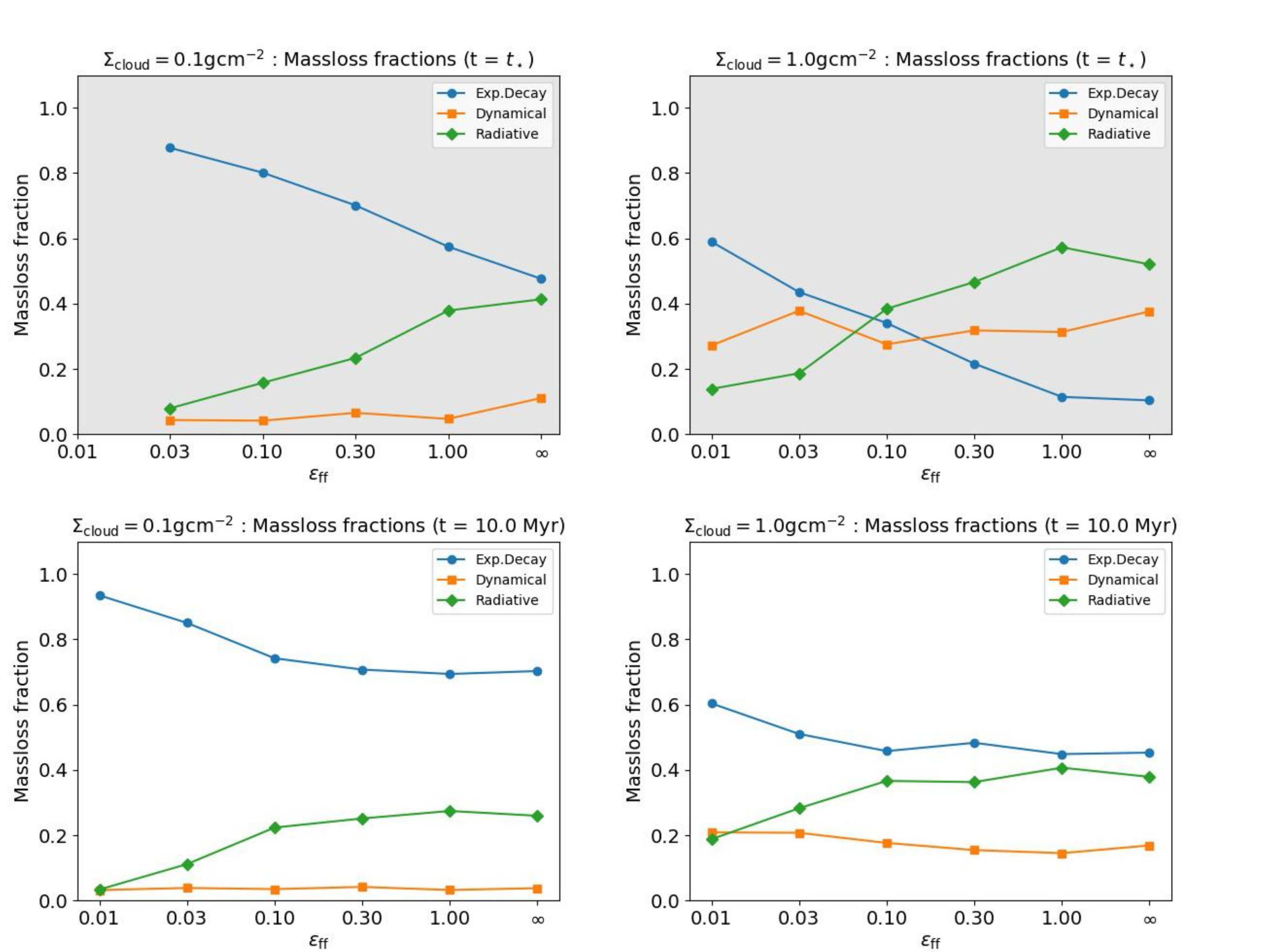}
    \caption{The fractional contribution to massloss by exponential decay (blue), dynamical truncations (orange) and external photo-evaporation (green) normalized to the total disc mass loss in the clusters at the time when star formation is complete (top row) and $t = 10.0$ Myr (bottom row) from the start of formation of the cluster.}
    \label{fig:fractionsstarformation}
\end{figure*}

In our models, internal evolution via exponential decay, binary truncations, dynamical truncations and external photo-evaporation all contribute to disc mass loss. In order to identify which mechanisms are dominant in different models ($\Sigma_{\rm cloud}$ and $\epsilon_{\rm ff}$), we studied the fractional contributions by these evolutionary mechanisms to the total disc mass loss. The top row of Figure~\ref{fig:fractionsstarformation} shows the contribution of different evolutionary mechanisms when star formation is complete, while the bottom row shows the contributions at the cluster age of 10 Myr. The left and right columns show the case for M3000L and M3000H sets, respectively. 

Exponential decay is the dominant mode of evolution of the discs. If we look at the top row, we observe that at the time when star formation is just complete, exponential decay contributes more to the disc mass loss for models that form more slowly (smaller $\epsilon_{\rm ff}$). This is because in the smaller $\epsilon_{\rm ff}$ models, exponential decay has had more time to operate on the discs by the corresponding $t_\star$.

At the time when star formation is just complete ($t=t_\star$), external photo-evaporation and dynamical truncations together have contributed up to a maximum of 58\% and 90\% of the total disc mass loss (in the \sfeff = $\infty$ model) in M3000L and M3000H sets, respectively. Between the two, external photo-evaporation is more dominant in the M3000L set. In contrast, discs in \sfeff = 0.01 and 0.03 models of the M3000H set have experienced more mass loss from dynamical truncations compared to external photo-evaporation. This is because the stellar densities are higher in the M3000H set, leading to more close encounters that truncate the discs. Moreover, there is less contribution from exponential decay because $t_\star$ is shorter in the M3000H models. 

As the clusters evolve, the relative contributions of the above mechanisms change. By $t$ = 10.0 Myr, external photo-evaporation is found to be dominant over dynamical truncations in both M3000L and M3000H sets. However, internal exponential decay still remains the main cause of disc massloss. The different $\epsilon_{\rm ff}$ models experience similar levels of exponential decay by this time. We should also mention the $\epsilon_{\rm ff}$ = 0.01 model in the M3000L set is still forming stars and thus is still embedded in natal gas at $t$ = 10.0 Myr. So the relative contribution of external photoevaporation appears to be smaller compared to the higher $\epsilon_{\rm ff}$ models. As soon as the natal gas is exhausted, external photoevaporation rises and dominates over dynamical truncations, as in the higher \sfeff models.

In this context, it is important to note that the different mechanisms compete with each other to drive disc mass loss. Both external photo-evaporation and dynamical truncations deplete the discs from outside-in, whereas exponential decay, as we have implemented it, only decreases the disc masses without changing the disc radii. Dynamical truncations could stop external photo-evaporation by truncating the disc within its gravitational radius $r_{\rm g}$, because the gas within $r_{\rm g}$ is tightly bound. Conversely, as photoevaporation reduces the disc radii, dynamical truncation events become less likely. When neither external photoevaporation nor dynamical truncations are significant, such as in a low density stellar environment where there are also no nearby massive stars or if discs are shielded by the natal gas, the discs can evolve relatively unperturbed (exponential decay in our models). For instance, these conditions are prevalent in the M3000L set before gas expulsion. 

\subsection{Dust mass distributions and comparison to the ONC and G286}
\label{subsec:dustmassdistros}

\begin{figure*}
    \includegraphics[width=\textwidth]{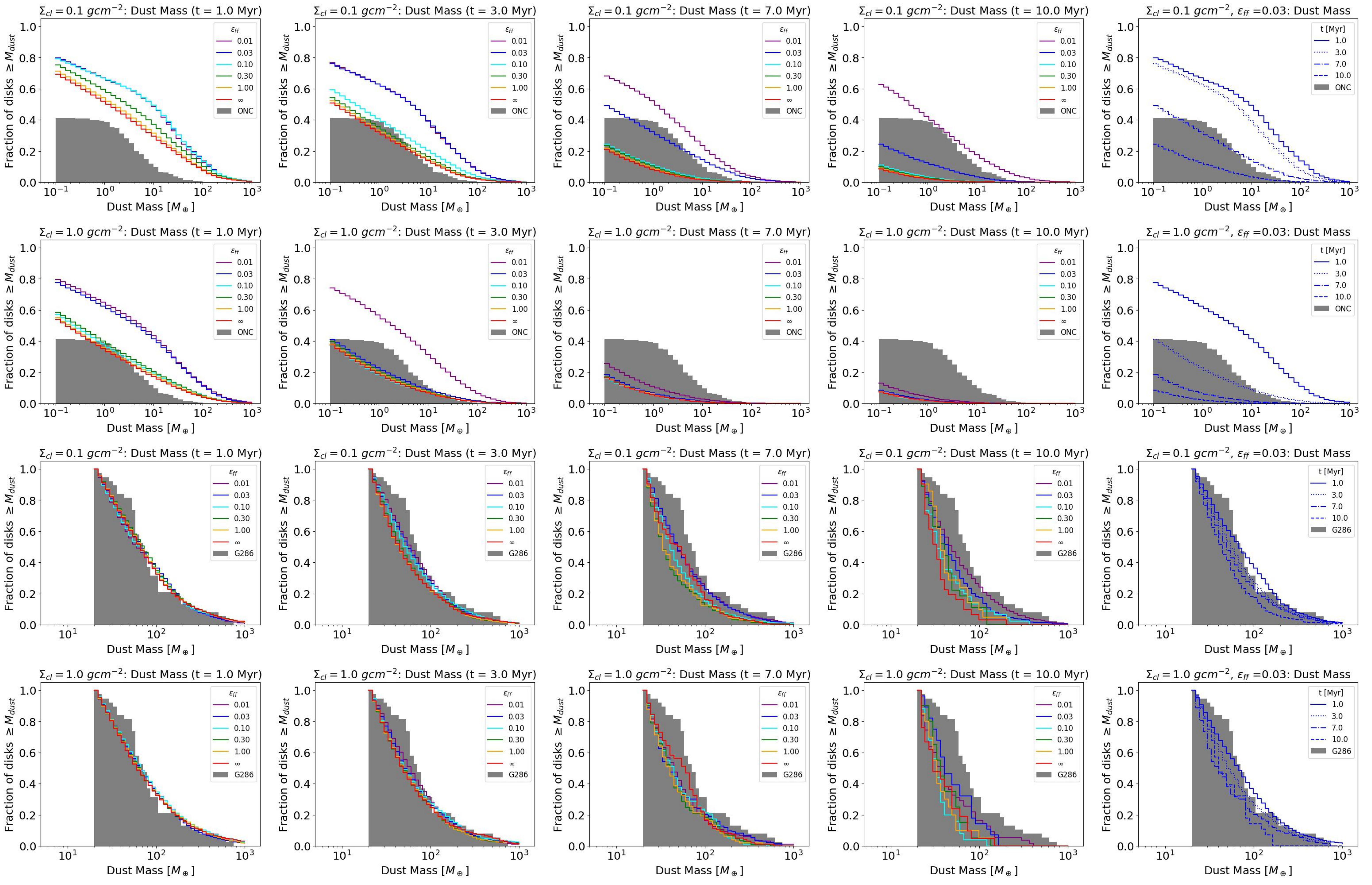}
    \caption{Comparison of dust masses in the ONC and G286 protoclusters with simulated dust mass distributions for M3000L clusters and M3000H clusters. In each panel, the y-axis shows the fraction of discs having dust masses equal or above the corresponding dust mass on x-axis. The first four columns show dust mass distributions at cluster ages of 1.0, 3.0, 7.0, and 10.0 Myr. The fifth column shows time evolution of dust mass for the fiducial $\sfeff = 0.03$ model for each row. The first two rows compare M3000L and M3000H sets with discs in the Orion Nebula Cluster. The next two rows compare M3000L and M3000H sets with discs in the G286 star-forming region.}
    \label{fig:dustmass_observations_comparison}
\end{figure*}

In order to study the effect of the cluster environment on the mass budget available for planet formation, we predict dust mass distributions in our simulations. The disc mass represents mass of gas in our simulated discs. Following \cite{Bohlin1978ApJ...224..132B}, we assume a dust to gas ratio of 0.01 to obtain the corresponding dust masses for each disc. We compared the dust masses across the different values of \sfeff and \Sigmacloud. Columns 1-4 of Figure~\ref{fig:dustmass_observations_comparison} show the dust mass distributions for the M3000L (first and third rows) and M3000H (second and fourth rows) sets at cluster ages of 1.0, 3.0, 7.0 and 10.0 Myr respectively. The fifth column shows the time evolution of dust mass distribution for the fiducial $\sfeff$=0.03 model in each set at $t$ = 1.0, 3.0, 7.0 and 10.0 Myr.

We observe that dust masses decrease with cluster age in all of our models, though at slightly different rates in different $\epsilon_{\rm ff}$ models. Once all stars are formed and the cluster ages, the dust masses in all models approach the instantaneous limit (\sfeff =$\infty$). In addition, from models \sfeff = 0.01 to $\infty$ at the same cluster age, there is a gradual decrease in fraction of discs having a certain dust mass. For the \sfeff = 0.01 and 0.03 clusters in M3000L set, the dust mass is larger, in agreement with the evolution of mean disc mass. If we consider the same \sfeff = 0.03 model at different times from 1.0 to 10.0 Myr in the fifth panel, we observe the gradual depletion in dust mass. The depletion of dust mass happens faster in M3000H set compared to the M3000L set. This is due to external photo-evaporation becoming more effective earlier in the M3000H set ($t_\star$ is shorter in M3000H set than corresponding M3000L set). 

In the same panels, we compare these dust mass distributions with observations of protoplanetary discs in two different star-forming regions observed with the Atacama Large Millimeter/Sub-millimeter Array (ALMA), as shown by the shaded histograms. The top two rows compare our models with Class II discs in the Orion Nebula Cluster \citep[ONC;][]{Eisner2018ApJ...860...77E} region. We compare our simulated discs to all sources, both detected discs and non-detections. We take logarithmic bins ($N = 40$) for dust mass from 0.1 $M_{\earth}$ to 1000 $M_{\earth}$. The leftmost bin is located at 0.1 $M_{\earth}$, which corresponds to the sensitivity limit of \citet{Eisner2018ApJ...860...77E}. For the ONC, 60\% of the sampled sources were found to be non-detections, i.e., below the sensitivity limit. We observe that different models better match ONC discs at different times. In the M3000L set, the fast-forming models (higher \sfeff) are more consistent with ONC discs at $t$ = 3.0 Myr. The slow-forming models (\sfeff = 0.03 and 0.01) are a closer match at 7.0 and 10.0 Myr, respectively. In the M3000H set, the fast-forming models are closest to ONC discs at 1.0 Myr, while the slowest-forming model (\sfeff = 0.01) has its closest match at 4.0 Myr.

Similarly, the next two rows compare our M3000L and M3000H models with discs in the G286 protocluster \citep{Cheng2022ApJ...940..124C}. The \citet{Cheng2022ApJ...940..124C} survey of G286 discs includes Class I/flat-spectrum sources with a sensitivity limit of 20 $M_{\earth}$. Since the study did not have statistics on the undetected discs below the sensitivity limit, we compare G286 with simulated discs above 20 $M_{\earth}$ only. Thus we take logarithmic bins ($N = 40$) for dust mass from 20 $M_{\earth}$ to 1000 $M_{\earth}$. We find that our \sfeff = 0.03 models best reproduce G286 disks around 1 Myr for both \Sigmacloud environments.


\subsection{Radial dependence of disc properties}
\label{subsec:radialdependence-results}

In the central dense core of a cluster, it is plausible to expect more stellar interactions leading to more dynamical truncations. Similarly, the more massive stars tend to segregate into the inner regions of the cluster by dynamical mass segregation \citep{Allison2009ApJ...700L..99A,Dominguez2017MNRAS.472..465D}. This could then lead to enhanced external photo-evaporation of discs in the cluster center. As a result, discs in the outer regions of the cluster may have larger masses and radii compared to the inner core where both dynamical and radiative feedback could be more active.

\begin{figure*}
    \centering
    \includegraphics[width=\textwidth]{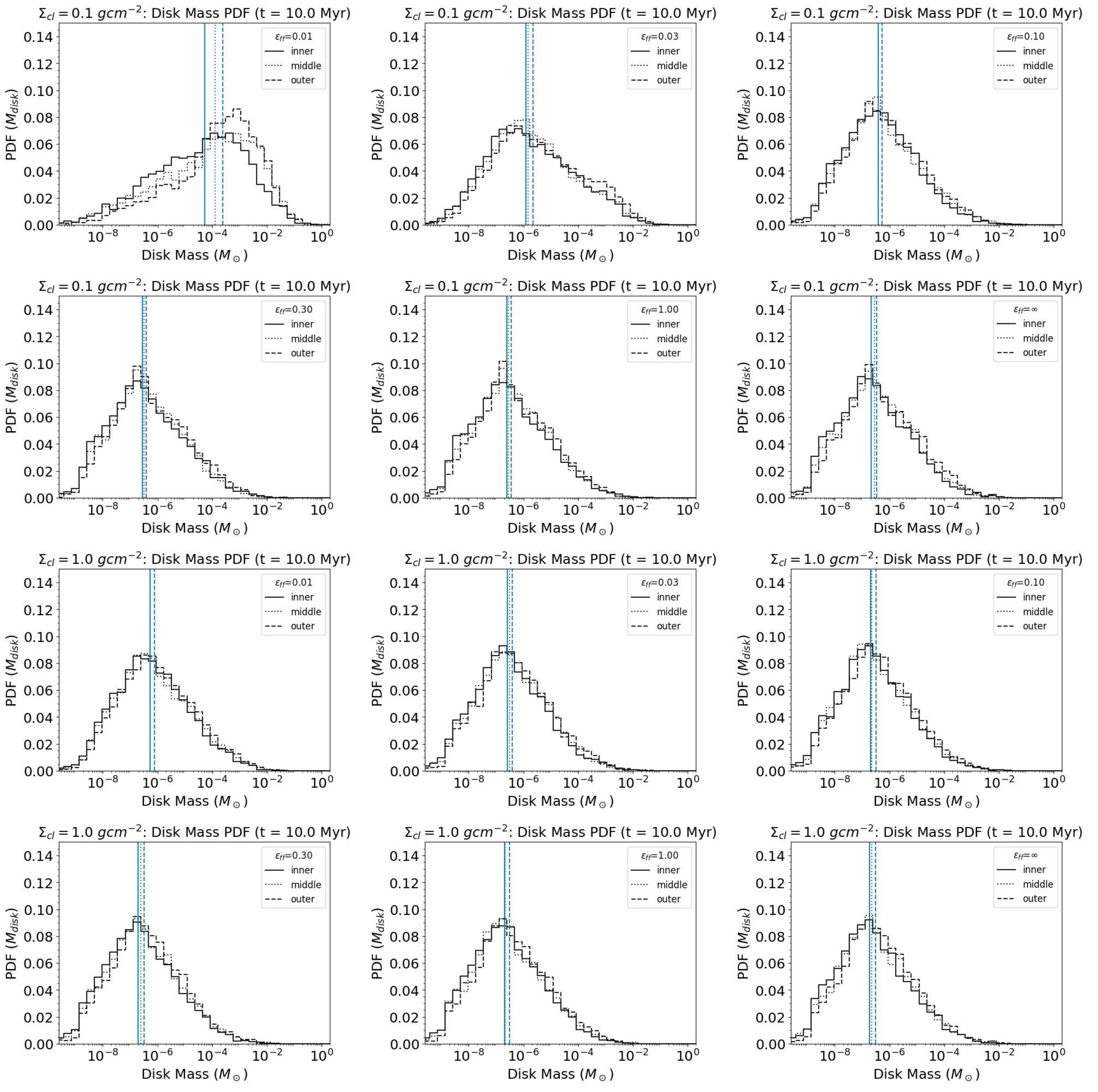}
    \caption{PDF of disc masses at 10 Myr around stars divided into three subsets (inner, middle and outer one-third) based on radial distance from the center of the cluster, shown for the different \sfeff models in the case of M3000L clusters (rows 1-2) and M3000H clusters (rows 3-4). The vertical solid, dotted and dashed lines show the median disc mass in these three subsets at 10 Myr.}
    \label{fig:radialprofilediskmass}
\end{figure*}

\begin{figure*}
    \centering
    \includegraphics[width=\textwidth]{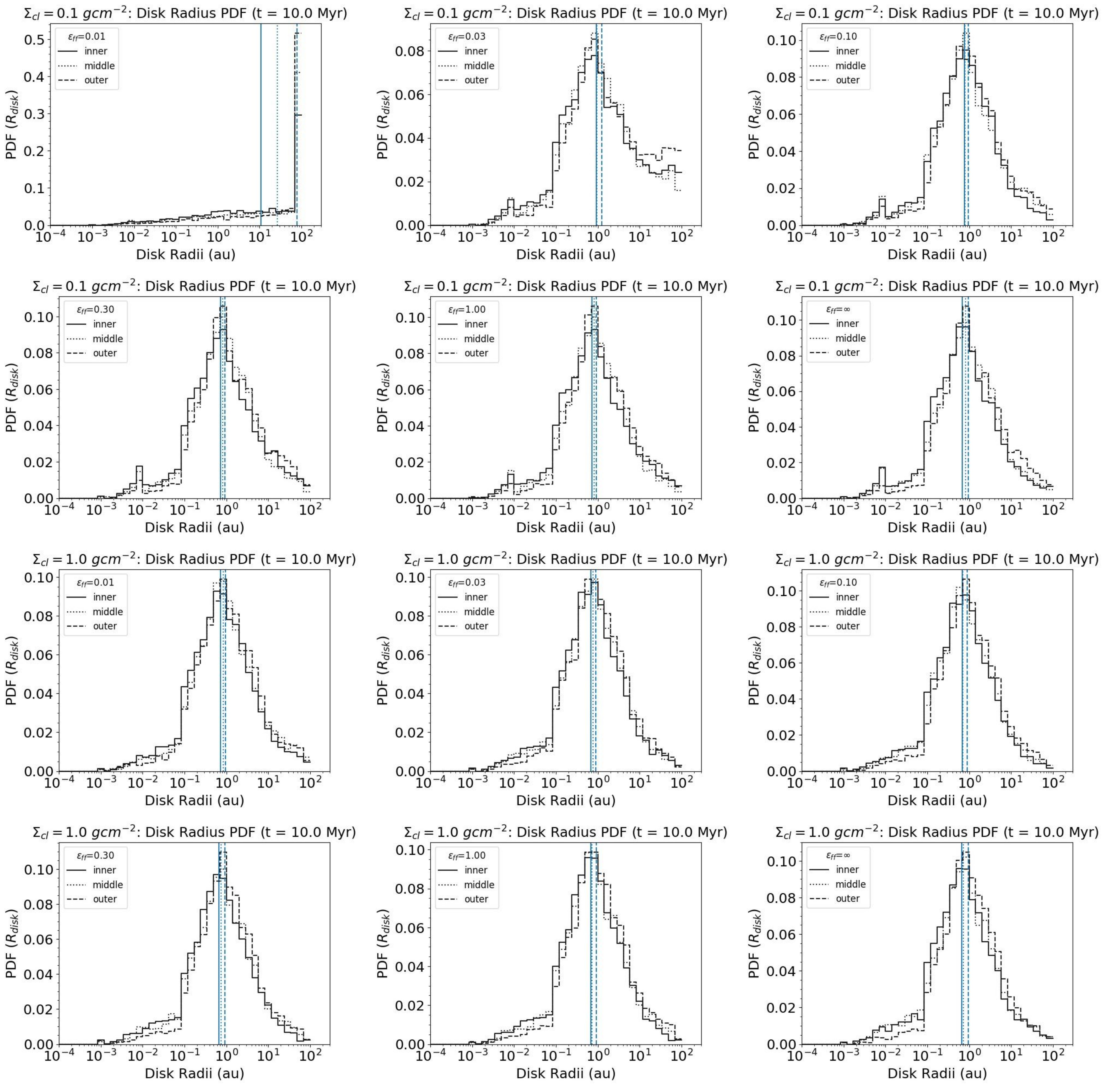}
    \caption{PDF of disc radii at 10 Myr around stars divided into three subsets (inner, middle and outer one-third) based on radial distance from the center of the cluster, shown for the different \sfeff models in the case of M3000L clusters (rows 1-2) and M3000H clusters (rows 3-4). The vertical solid, dotted and dashed lines show the median disc radius in these three subsets at 10 Myr.}
    \label{fig:radialprofilediskradius}
\end{figure*}

To test this hypothesis, we studied the variation of the median disc mass and median disc radius, between the inner and outer regions of the cluster. We divided discs into three subsets based on their projected (2D) radial distance from the center of the cluster: inner, middle and outer. Figures \ref{fig:radialprofilediskmass} and \ref{fig:radialprofilediskradius} show the median disc mass and median disc radius, respectively, in these subsets along with their PDFs. We observe that median disc mass and median disc radii of the inner subset are always the smallest among the three subsets across all $\Sigmacloud, \sfeff$ models. In other words, the discs in the outer regions of the cluster are larger and more massive than discs in the inner regions. For the fiducial \sfeff = 0.03 case, the median disc in the inner center is approximately 50\% and 35\% less massive than in the outer regions for M3000L and M3000H sets, respectively. The differences between inner and outer subset are most prominent in the \sfeff = 0.01 model of M3000L set, with inner discs being 77\% less massive than outer discs. We checked whether discs are larger than their gravitational radii ($r_g$) at $t$ = 10 Myr, given by equation \ref{eqn:gravitationalradius}. We found that most of the discs in the inner core are smaller than their corresponding $r_{g}$ and thus are not susceptible to photo-evaporation anymore. In addition, their small size (around 1 AU) means that dynamical truncations are also less likely. In other words, the cluster environment has already shaped these discs to a large extent. In contrast, discs larger than $r_g$ were preferentially located on the outer regions of the cluster. We discuss previous observations of radial trends in disc properties in section~\ref{subsec:radialdependence-discussion}.


\section{Discussion}
\label{sec:discussion}
We have studied the evolution of protoplanetary discs inside gradually forming star clusters. These star clusters were formed in different star formation environments determined by star formation rates (\sfeff) and mass surface densities (\Sigmacloud) of the parent cloud. Inside each star cluster, the discs evolved under the influence of external photo-evaporation and dynamical truncations, in addition to a general exponential decay of disc mass with time.  

\subsection{Radiation shielding of discs by natal gas}
\label{subsec:shielding}
We observed that external photoevaporation has a dominant influence on the evolution of discs after star formation is complete. In this regard, radiation shielding by the background gas is important in protecting discs from external photoevaporation before star formation is complete. This helps them retain larger dust masses for longer times compared to the models in which the ionizing radiation passes unimpeded. This has been noted previously by \cite{Qiao2022MNRAS.512.3788Q} who found discs were shielded by the natal gas for up to 0.5 Myr after the start of star formation in their simulations. \cite{Wilhelm2023arXiv230203721W} also found that disc lifetimes and mass reservoirs were increased by radiation shielding. The timescales on which disc mass is lost may be important for planet formation. If the gas and dust in discs dissipate quickly, then there would be a limited time-frame in which planets could form \citep[see][]{Qiao2023MNRAS.522.1939Q}. We also observe that the timescale of shielding provided by the natal gas depends on \sfeff and \Sigmacloud in our models because these determine the timescale of cluster formation and expulsion of gas. For instance, discs in the slower-forming \sfeff = 0.01 and 0.03 models are shielded for longer in general than discs in fast-forming models. This is in agreement with larger disc masses and disc radii observed for \sfeff = 0.01 and 0.03 discs at early cluster ages/ disc ages in Figures \ref{fig:meandiskpropertyevolution_clusterage} and \ref{fig:meandiskpropertyevolution_diskage}, respectively. In addition, the extent and duration of shielding for individual discs could be determined by their proximity of formation to and/or encounters with massive stars. This depends on the dynamical state of the cluster during the embedded phase when the discs are formed.  

\subsection{Radial dependence of disc properties}
\label{subsec:radialdependence-discussion}

In section~\ref{subsec:radialdependence-results}, we observed the global trend of discs in the outer regions of a cluster being larger and more massive than the discs in the inner core. Previous observational studies have reported positive correlations at a local level between the mass of a photo-evaporating disc and distance from a nearby massive star in star-forming regions like the ONC \citep{Mann2014ApJ...784...82M} and $\sigma$ Orionis \citep{Ansdell2017AJ....153..240A}. In these regions, discs closer to the most massive star were found to be less massive and smaller in size than distant discs. However, \cite{Mann2015ApJ...802...77M} did not find such correlations in NGC 2024. Moreover, \cite{Parker2021ApJ...913...95P} found that such correlations could be created by projection effects in observations. When mapping the 3D structure of the cluster to 2D projected distances, we may confuse foreground or background discs to be closer to the massive star and thus infer inaccurate correlations. Finally, cluster dynamics could move discs closer to and further from massive stars at different times, mixing away any local correlations. 

Our results, however, concern the global trend of disc properties from the center of the cluster. Even though discs move around the cluster with their host stars and smear out any local correlations around a particular massive star, we found the global trend of having larger discs in outer regions remains apparent in the clusters. If dynamical segregation of massive stars to the center occurs, the discs present in outer regions during expansion of the cluster would be left relatively untouched by the ionizing radiation field in the core. It is important to note that we examined the global trend after the formation of the cluster. During ongoing star formation, the natal gas tends to attenuate the radiation fields caused by segregated massive stars even more strongly for discs in the outer regions and leads to more visible differences in inner and outer discs, as observed in M3000L, \sfeff = 0.01 model at 10 Myr ($t_*$ = 20 Myr). 

\subsection{Model caveats}
\label{subsec:modelcaveats}

Our aim has been to carry out a first exploration of disc evolution in the context of gradually forming clusters to identify which processes affect discs the most in different star formation environments. To make this first investigation of disc evolution relatively simple and reduce the number of input parameters, we have assumed a fixed initial ratio of disc mass to host mass (although this is also modified by primordial binary truncation) and a fixed initial disc size of 100~AU. Future work in this series will examine the impact of various distributions in these inputs.

We have also considered relatively simple models for the evolution of discs in our stellar clusters, in part motivated by the need for these to be implemented efficiently in our $N$-body code and utilizing input parameters that are captured in these simulations. For example, in modeling of dynamical truncations, we have not considered the exact orientations of the discs during stellar encounters. In encounters between two disc-hosting stars, we also do not consider the interactions between the discs themselves, or the tidal effects of the discs on stellar dynamics. We model only the effect of the perturbing star upon the host disc using equation \ref{eqn:breslau}. Similarly, the potential accretion of truncated disc material onto the host or perturbing star was not considered. The accretion of truncated disc material onto the host or perturber during stellar encounters is a possible explanation for outburst events observed in FU Orionis objects \citep{Cuello2022arXiv220709752C}. In these objects, extreme accretion events driven either by stellar flybys \citep{Borchert2022MNRAS.517.4436B} or bound binary interactions \citep{Bonnell1992ApJ...401L..31B} are posited to trigger large increase in luminosities within a short timescale of years. 

Another important caveat is that we have assumed a smooth background gas potential in our simulations. In reality, the distribution of gas in star forming regions is more complex with substructures like bubbles and filaments \citep{Hacar2022arXiv220309562H}. In particular, \cite{Wilhelm2023arXiv230203721W} noted that sub-structured cluster gas could create very large spatial gradients in the radiation field experienced by a disc. There could be regions relatively rarefied in gas, where discs are especially susceptible to photoevaporation. This means shielding provided by the structured gas can be less pronounced than the case of a smooth distribution. In such a case, external photo-evaporation could have a greater influence during the embedded star forming phase as well, not just after the dissipation of cluster gas.   

We have not considered supernova feedback in our models. The number and timing of supernovae depends on the sampling of the IMF. On average, we find around 3 supernovae happen per model realization by 10 Myr for M3000L set and 4 supernovae per realization for M3000H set (with this number being a little greater on average because star formation finishes earlier). A future project in this series will examine a wider range of cluster conditions, including a range of cluster masses, and include study of the rates of supernovae and their proximity to discs providing information about enrichment of discs with heavier elements formed during supernovae \citep{Williams2007ApJ...663L..33W,Gounelle2008ApJ...680..781G}.

\section{Conclusions}
Planet formation around a star could be shaped from an early stage by the environment of its star cluster. We have implemented simple models of protoplanetary disc evolution within $N$-body models of gradual star cluster formation. Modeling the evolution of the discs during the star cluster formation simulation itself, we have studied the relative importance of different mechanisms acting on the discs. This includes external photoevaporation due to massive stars (radiative feedback) and disc truncations due to stellar encounters (dynamical feedback). We also take into account the effects of primordial binary partners, as well as considering a general exponential decay for all discs in isolation. 

We found that the rate of star formation (\sfeff) and the density of parent cloud (\Sigmacloud), which control the formation timescale of the star cluster, influence the evolution of protoplanetary discs in a forming star cluster. Radiative and dynamical feedback from the cluster can reduce dust masses in the discs with implications for planet formation. In this regard, external photoevaporation due to massive stars is found to be more important than dynamical encounters across the range of initial conditions (\sfeff and \Sigmacloud) we consider. As such, shielding provided by the background gas against the ionizing radiation fields is found to be important in determining dust-masses and lifetime of discs. discs in the slow-forming clusters are shielded for longer than fast-forming clusters, and thus provide a longer window for planet formation to occur. 

Next, we compared our discs with observations in the Class II discs in the ONC and Class I/Flat-Spectrum discs in G286 star-forming regions, respectively. Our discs in the fiducial \sfeff = 0.03 model resemble ONC discs around 3 Myr and 7 Myr for low and high \Sigmacloud, respectively. Meanwhile, our discs in the fiducial \sfeff = 0.03 model resemble G286 discs around 1 Myr. Finally, we also observe that discs in the inner core of clusters are less massive and smaller than discs in the outer regions, hinting at disc depletion mechanisms being more active there due to mass segregation of massive stars and/or shielding of discs in outer regions. 

\section*{Acknowledgements}
AG acknowledges invaluable support from the Chalmers Astrophysics and Space Sciences Summer (CASSUM) research program. JCT acknowledges support from ERC Advanced Grant 788829 (MSTAR). The simulations presented in this work were performed on the Origins computing cluster at Chalmers. This research has made use of NASA’s Astrophysics Data System.  

\section*{Data Availability}
The data underlying this article will be shared on reasonable request to the corresponding author.



\bibliographystyle{mnras}
\bibliography{references} 








\bsp	
\label{lastpage}
\end{document}